\documentclass[twocolumn,english,showpacs,amsmath,amssymb,aps,prl,floats]{revtex4}
\usepackage[]{fontenc}
\usepackage[latin1]{inputenc}
\usepackage{subfigure}
\usepackage{amsmath}
\usepackage{color}

\usepackage{epsfig}

\usepackage{dcolumn}

\usepackage{bm}

\usepackage{amsfonts}

\usepackage{graphicx}%

\newcommand{\bra}[1]{\langle #1 |}
\newcommand{\ket}[1]{| #1 \rangle}
\newcommand{\be}{\begin{equation}}
\newcommand{\ee}{\end{equation}}
\newcommand{\beq}{\begin{eqnarray}}
\newcommand{\eeq}{\end{eqnarray}}

\usepackage{babel}
\makeatother
\begin{document}

\title{Phase separation in fermionic systems with particle-hole asymmetry}

\author{Arianna Montorsi}
\affiliation{Dipartimento di Fisica del Politecnico and CNISM, corso Duca 
degli Abruzzi 24, I-10129, Torino, Italy}
\date{March 12, 2008}

\begin{abstract} We determine the ground-state phase-diagram of a Hubbard Hamiltonian with correlated hopping, which is asymmetric under particle-hole transform. By lowering the repulsive Coulomb interaction $U$ at appropriate filling and interaction parameters, the ground state separates into a hole and an electron conducting phases: two different wave vectors characterize the system and charge-charge correlations become incommensurate. By further decreasing $U$ another transition occurs at which the hole conducting region becomes insulating, and conventional phase separation takes place. Finally, for negative $U$ the whole system eventually becomes a paired insulator. It is speculated that such behavior could be at the origin of the incommensurate superconducting phase recently discovered in the 1D Hirsch model. The exact phase boundaries are calculated in one dimension. 

 \end{abstract}

\maketitle

One very debated issue in the context of correlated fermionic materials is the occurrence of   phase separation (PS). For instance, in ultracold fermionic atoms trapped on optical lattices PS appears when the superfluid phase coexists with the normal phase \cite{PAR,PIGI}; these systems may be investigated by means of the negative $U$ Hubbard model \cite{XIal}. Also, in high-$T_c$ materials PS leads to the formation of one-dimensional stripes\cite{TRANQ}. It is believed that the phenomenon could be a consequence of the interplay between antiferromagnetic order and holes propagation; as such, its appearance is investigated within the context of the $t-J$ model \cite{EKL,PLR,HEMA,CBS}, which describes the Hubbard model in the limit of strong repulsive Coulomb interaction. In the present paper we suggest a further scenario for the occurrence of PS in Hubbard-like systems, resulting from competition between conducting phases with different charge: in the hole rich region the carriers are the electrons, while in the pair rich region they eventually become the holes. In absence of particle-hole symmetry, the proposed mechanism can generate incommensurability. The mechanism is proven to survive in dimension greater than 1 for a simple case. 

We study the bond-charge extended Hubbard Hamiltonian, which has been introduced to describe compounds containing extended orbitals \cite{cam}. It comes from assuming that the charge in the bond affects the effective potential acting on valence electrons; the extension of the Wannier orbitals and the hopping between them should vary with the charge. A generalization of that Hamiltonian is the $U-X-X'$ Hamiltonian introduced in \cite{SA}. In the grand-canonical ensemble it reads:
\begin{widetext}
\be
H_{BC} =  -\sum_{\sigma=\uparrow,\downarrow,\langle ij\rangle}
\left [1-x(n_{i-\sigma} +n_{j-\sigma})+x' n_{i-\sigma}n_{j-\sigma}\right ] 
(c_{i\sigma}^{\dagger}c_{j\sigma}+{\text H.c.})\nonumber + 
u\sum_{i} n_{i\uparrow}n_{i\downarrow}-\mu \bigl (\sum_{i\sigma}n_{i\sigma}-N \bigr )  
\quad , \label{ham_BC}
\ee
\end{widetext}
where the lower case symbols denote that the coefficients of the interactions have been normalized in units of the hopping amplitude, and the 3-body interaction term $x'$ appears as an effective interaction from the three band-model. Moreover $N$ is the number of electrons on the $D$-dimensional $L$ sites lattice $\Lambda_D$. The model belongs to the class of Hubbard Hamiltonians with correlated hopping, which contains some integrable cases that have been widely studied in recent years (see \cite{corrhop} and references therein).

$H_{BC}$ is in general is not invariant under particle-hole transform. For this reason the model has been first proposed (for $x' =0$) in two dimensions, motivated by a theory of hole superconductivity \cite{hirsch}. It was investigated by means of bosonization approach in \cite{JAKA} at half-filling for arbitrary (and weak) $x$ and $x'$.  Also, a particle-hole invariant subcase of $H_{BC}$, which takes place at $x'=2 x$, has been previously discussed\cite{NAKA}. 
\\Recently, it has been proven \cite{ADMO},\cite{AAA} that in 1D at half-filling $H_{BC}$  displays a rich phase diagram already at $x'=0$, quite different from that of the Hubbard model, and from that derived in the weak coupling limit \cite{JAKA}. In particular, at sufficiently low $u>0$ , and for appropriate range of $x$ values an unexpected incommensurate superconducting (ICSS) phase appears, which drives the usual transition to an insulating state to higher values of $u$. Here we investigate the origin of such phase by constructing the exact ground-state phase-diagram of a particular sub-case of (1) still asymmetric under particle-hole transform.

\begin{figure}
\includegraphics[width=80mm,keepaspectratio,clip]{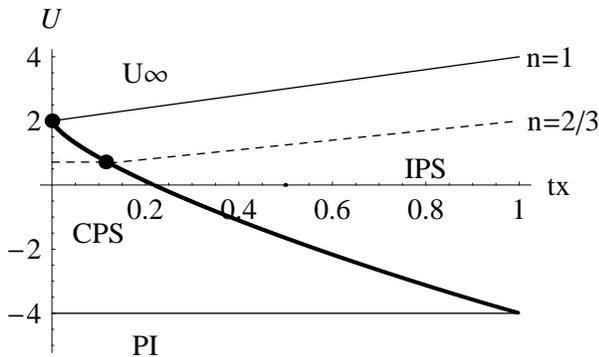}
\caption{(Color online) Phase diagram in the $u$-$t_x$ plane in $D=1$: the transition to the $U\infty$ region depends on the filling $n$, whereas the other transitions are independent of it. CPS and IPS denote the conventional and the incommensurate phase separation regions respectively; PI stays for paired insulator. Full dots indicate the tricritical points.}  \label{fig1}
\end{figure}

Hamiltonian (\ref{ham_BC}) can be written in terms of the Hubbard projectors, 
$X^{\alpha\beta}_i\doteq\ket{\alpha}_i\bra{\beta}_i$; here $\ket{\alpha}_i$ 
are the states allowed at a given site $i$, $\alpha=0,\uparrow,\downarrow,2$ ($\ket{2}\equiv 
\ket{\uparrow\downarrow}$). Inserting the choice $x=1$ in $H_{BC}$, the resulting Hamiltonian $H\equiv H_{BC} (x=1)$ turns out to preserve the number of doubly occupied states $N_d= < \hat N_d> \doteq <\sum_j X^{22}_j>$: $[H,\hat N_d]=0$ for arbitrary $x'$, $D$ and $N$. In the following, we shall adopt such choice. $H$ can then be written as
 $$H=H_{01}+H_{12}-\mu(L-N) \quad , \label{ham}$$
where
\begin{eqnarray}
H_{01} &=& - \sum_{\langle i,j\rangle \sigma } 
(X_i^{\sigma 0}X_j^{0\sigma}+h.c.)+ \mu \sum_i X_i^{00}\\
H_{12} &=&  - t_x \sum_{\langle i,j\rangle \sigma} (X_i^{2\sigma}X_j^{\sigma 2}+ 
h.c.)+ (U-\mu) \sum_i 
X_i^{22} \; , \label{Hpieces}
\end{eqnarray}
with $t_x=1-x'$. We shall limit our analysis to the range $0\leq t_x\leq 1$. By implementing the transformation $c_{i,\sigma}\rightarrow c_{i,\sigma}^\dagger$, it can be realized that the range is representative of the behavior of the model at any $t_x$ value. In the two limits $t_x=0$ ($x'= 1$) and $t_x=1$ ($x'=0$) the Hamiltonian reduces to known cases, namely the infinite $U$ Hubbard model, and the bond-charge Hubbard model at $x=1$ \cite{AAS}, both integrable in one dimension. 
\\On general grounds, we expect that at large enough positive $u$ the ground state would contain the minimum number of doubly occupied sites. Hence it will coincide with that 
of the infinite $U$ Hubbard model (no doubly occupied sites) for $N\leq L$, and with its particle-hole counterpart 
for $N\geq L$. In Fig. \ref{fig1} such state is denoted as $U\infty$.  Also, 
for large enough negative $u$, the number of doubly occupied sites will be 
maximized, so that for even number of electrons the hopping term will be ineffective and the ground state would be a highly degenerate insulator consisting of $N/2$ pairs of electrons. In Fig. \ref{fig1} this is denoted as $PI$ (paired insulator).  
In order to investigate what happens between these two limits, let us think of  
the Hilbert space as factorized into three orthogonal subspaces, 
\begin{equation}
{\cal{H}}= {\cal{H}}_{01}\oplus {\cal{H}}_{12}\oplus {\cal{H}}_\perp \quad :
\end{equation} 
states with no doubly occupied sites belong to ${\cal{H}}_{01}$, 
states with no empty sites live in ${\cal{H}}_{12}$, and the 
remaining states stay in ${\cal{H}}_\perp$. $H_{10}$ does not act on doubly occupied sites present in ${\cal{H}}_\perp$, and annihilates states belonging to ${\cal{H}}_{12}$; hence, at given $N_d$ and $N$, it reaches its minimum in ${\cal{H}}_{01}$. Analogously,  $H_{12}$ does not act on empty sites present in ${\cal{H}}_\perp$, and annihilates states in ${\cal{H}}_{01}$, hence reaching its minimum in ${\cal{H}}_{12}$. So that, at any given $N$ and $N_d$, the absolute minima of both $H_{10}$ and $H_{12}$ (and consequently of $H$) are reached in the space  orthogonal to ${\cal{H}}_\perp$. We can then rewrite the ground state energy $E_{gs}$ in the form
\begin{equation}
E_{gs}= min_{\ket{\psi} \in {\cal{H}}} \bra{\psi} H\ket{\psi} \equiv 
min_{\ket{\psi} \in {\cal{H}}_{01}\oplus {\cal{H}}_{12}} 
\bra{\psi} H\ket{\psi}  
\quad ,
\end{equation}
where the minimum has to be taken with respect to $N_d$, 
the constraint $L-N=N_e-N_d$ is implemented through the chemical potential, and $N_e\doteq \sum_i X_i^{00}$ is the (conserved) number of empty sites. More 
explicitly, one could look for a ground state of the form:
\begin{equation}
\ket{\psi(N_e,N_d)}_{gs} = a\ket{\psi_{01}(N_e')}+\sqrt{1-a^2} \ket{\psi_{12}(N_d')}\quad ,  \label{gs}
\end{equation}
where  $a$ is a further variational parameter, and $N_e= a^2 N_e '$, $N_d= (1-a^2) N_d'$. Here $\ket{\psi_{01}(N_e')} \in 
{\cal{H}}_{01}$, and $\ket{\psi_{12}(N_d')}\in {\cal{H}}_{12}$ are the ground 
states of $H_{01}$ at given $N_e'$ and $H_{12}$ at given $N_d'$ respectively. The corresponding energies are recovered from those of the infinite $U$ Hubbard model, $E_{gs}^{\infty}$. In fact, the spectra of $H_{01}$ and $H_{12}$ are the same: up to a multiplicative factor $t_x$, and to the conserved quantities $N_e$, $N_d$, they coincide with the spectrum of the infinite $U$ Hubbard model. Explicitly, 
\beq
E_{gs} &=& a^2  E_{gs}^{\infty} (N_e') + 
(1- a^2)\left [ 
t_x E_{gs}^{\infty} (N_d') + u N_d' \right]\nonumber \\ &-&\mu 
\left (N_d-N_e+L-N \right ) 
\quad . \label{Egs}
\eeq
By minimization 
of $E_{gs}$ with respect to $\mu$ one obtains that the actual value of $a^2$ 
is given by 
\be
 a^2= \frac{L-N+N_d'}{N_e'+N_d'} \quad .
\ee
The case $a^2=1$ ($a^2=0$) correspond to the $U\infty$ 
phase described above for $N\leq L$ ($N\geq L$), with $N_e'= L-N$ ($N_d'=N-L$).
Apart from these limits,  the system is always separated into two phases, 
characterized by $N_e'$, $N_d'$, and $\mu$ values which minimize $E_{gs}$ 
as given by (\ref{Egs}), at given $t_x$ and $u$. Such values do not 
depend on the actual filling $N$: PS implies that the 
chemical potential is constant with the filling. 
\\Since $t_x\leq 1$, independently of the actual value of $E_{gs}^{\infty}$ (and $D$)
one could expect three different behavior to emerge from the minimization 
equations within the PS phase, depending on the filling and the 
interaction parameters.
\begin{description}
\item[(i)] $N_e'< N_d'< L$: both the coexisting phases are conducting. They are characterized  by different Fermi momenta $k_F^{(e)}$ and $k_F^{(d)}$, which can be inferred from those of the corresponding infinite $U$ model, and are in general not commensurate between each other. For this reason we characterized the phase as IPS, incommensurate phase 
separation. Later on we shall discuss how such incommensurability reflects 
onto some physical feature. 
\item[(ii)] $N_e'< N_d'=L$; only the hole-rich region is conducting, 
whereas the other phase is a paired insulator. The system is characterized by the Fermi
momentum $k_F^{(e)}$ of the conducting electrons, {\it i.e.} that of the infinite $U$ Hubbard model at a filling $L-N_e'<N$. We denote such phase as $CPS$ (conventional phase separation). 
\item[(iii)] $N_e'=N_d'=L$; both phases are insulating. In this case we are 
in one of the infinitely many (in the thermodynamic limit) degenerate ground states 
characterizing the PI phase. 
\end{description}  
A further interesting limiting case, not discussed above though  contained in 
the minimization equations, occurs when $N_e'=N_d'\neq L$. In such case both phases are conducting as in (i), but they are characterized 
by the same Fermi momentum, $k_F^{(e)}=k_F^{(d)}$. This is the case for $t_x=1$. 
The latter is the already mentioned bond-charge Hubbard model at $x=1$ \cite{AAS}, 
and manifests particle-hole invariance. This example illustrates how in our scheme the IPS 
phase, intended as the appearance of two not commensurate modulations, is in fact 
associated with the presence of particle-hole asymmetry.  

In order to make the above predictions more quantitative, we now turn to the 
discussion of the 1D case. In that case the ground-state energy per-site of 
the infinite $u$ Hubbard model is known exactly. In the thermodynamic limit 
it reads: 
\be
E^\infty_{gs}(n) =-\frac{2}{\pi} \sin (\pi n) \quad , 
\ee
with $n$ average per-site filling, $n\doteq \frac{N}{L}$.\\ 
When inserted into $E_{gs}/L$, as given by (\ref{Egs}), the above information 
allows for an explicit derivation of the ground-state phase diagram, shown in Fig. \ref{fig1}. 
This is achieved by solving the minimization equations with respect to  $n_e'\doteq {N_e'\over L}$ and $n_d'\doteq {N_d'\over L}$, which in turn fix $k_F^{(e)}=\pi (1-n_e')$ and $k_F^{(d)}=\pi (1-n_d')$. \\ Interestingly, the same phase diagram and ground-state energy can be also obtained by generalizing the approach developed in \cite{PEMI} for studying PS in the extended Hubbard model for $V\rightarrow \infty$, with $V$ neighboring sites diagonal Coulomb interaction. At a given filling, the ground state can be thought of as an island of $N_d+Z$ singly and doubly occupied sites inserted into a chain of $L-N_d-Z$ singly and empty sites. Such choice allows to maximize the number of available low energy momenta. In the thermodynamic limit, minimizing the energy of that state with respect to $N_d$ and $Z$ gives precisely the same phase diagram obtained here in Fig. 1. It turns out that the following relations hold: $n_d'=\frac{N_d}{N_d+Z}$, and $n_e'=\frac{L-N+N_d}{L-N_d+Z}$.

The solutions $n_e'=1-n$ ($a^2=1$) below half-filling, and $n_d'=1$ ($a^2=0$) above half-filling, are obtained for $u>u_c(n,t_x)$ (see below). These identify the  $U\infty$ phase. In particular, at $n=1$ a charge gap $\Delta_c$ opens and the phase becomes insulating: for $u\ge u_c (1,t_x)= 2 (t_x+1)$, $\Delta_c=u-u_c$. 
\\For $a^2\neq 0,1$ the solution to the minimization equation reads:
\be
n_e'= \frac{1}{\pi} \arccos \left (\frac{\mu}{2}\right ) \; , \; n_d'= \left \{\begin{array}{ll}
\frac{1}{\pi} \arccos \left (\frac{u-\mu}{2 t_x}\right ) &\mbox{IPS}\\
1 &\mbox{CPS}
\end{array}\right. \quad . \label{xy}
\ee
Minimizing $E_{gs}$ is thus reduced to solving the following transcendental equation in $\mu$, 
\be
\mu= \frac{1}{n_e'+n_d'} \left ( - 2 t_x\sin {\pi n_d'}+ u n_d'+2 \sin 
{\pi n_e'} \right ) \quad , \label{mueq}
\ee
with $n_e'$ and $n_d'$ as specified in (\ref{xy}). Let us denote the solution by 
$\bar \mu (u,t_x)$. The actual value of $\bar \mu_I$ 
in the IPS phases  is limited by the constraint 
\be
\bar \mu_I \in \left ] \max  (-2, u-2 t_x), \min (2, u+2 t_x)\right [ \quad .
\ee
Whereas in the CPS phase $\mu$ becomes independent of $t_x$, and $\bar\mu_{C}(u)$ is the solution of (\ref{mueq}) for $n_d'=1$. In fact,  all the transition curves can be characterized by the value of $\bar\mu$ along them:
\begin{enumerate}
\item{} $U\infty \rightarrow IPS, CPS$, 
\be \bar\mu (u,t_x) = 
\left \{\begin{array}{ll}
- 2 \cos (\pi n) &n\leq 1\\
u+ 2 t_x \cos (\pi n) &n\geq 1
\end{array}\right. \label{ucr}
\ee
\item{} $IPS\rightarrow CPS$, $\bar \mu_I=\bar\mu_{C}$; the transition occurs at $\bar t_x=[\bar\mu_C (u)-u]/2$, implying $n_d'=1$. 
\item{} finally at $CPS\rightarrow PI$ $\bar\mu_{C}=-2$.
\end{enumerate}
When passing from $\mu$ to $n$, 
one realizes that only the two transitions $1$ do depend on the filling, whereas the others are independent of it. At transition $1$ the explicit value of $u_c(n,t_x)$, is obtained by solving eq. (\ref{ucr}) for $u$. Transition $2$ takes place at $t_x=\bar t_x$; transition $3$
occurs at $u=-4$. As an example, we report the critical curves in fig. \ref{fig1}  in the ($u,t_x$) plane at two different filling values. For $u\leq 2$, depending on $n$, two tricritical point (T1) are recognized, characterized by the merging of $U\infty$, IPS and CPS phases: one (not shown) occurs at $n=2$, whereas the other one occurs at the same $u$ value and at a filling $n_{T1}\le 1$ . Analytical calculations show that $u_{T1}= - 2 (\cos [\pi n_{T1}]+t_x)$, $n_{T1}\approx {1\over \pi} \arccos [(3 t_x-1)/(1+t_x)]$. Increasing $t_x$ progressively drives the system to a particle-hole symmetric diagram. For $t_x=1$ $n_{T1}=0$, which is the particle-hole counterpart of the tricritical point in $n=2$.  The critical curves in the ($u$,$n$) plane (not shown) highlight that particle-hole asymmetry ($t_x\neq 1$) favors the IPS phase at filling greater than half, in which case it survives at appropriate positive $u$ values up to $n=2$.  Fig. \ref{fig1} also shows that the infinite $U$ Hubbard model ($t_x=0$) exhibits CPS for $u\le 2$ if doubly occupied sites are allowed. 

\begin{figure}
\includegraphics[width=70mm,keepaspectratio,clip]{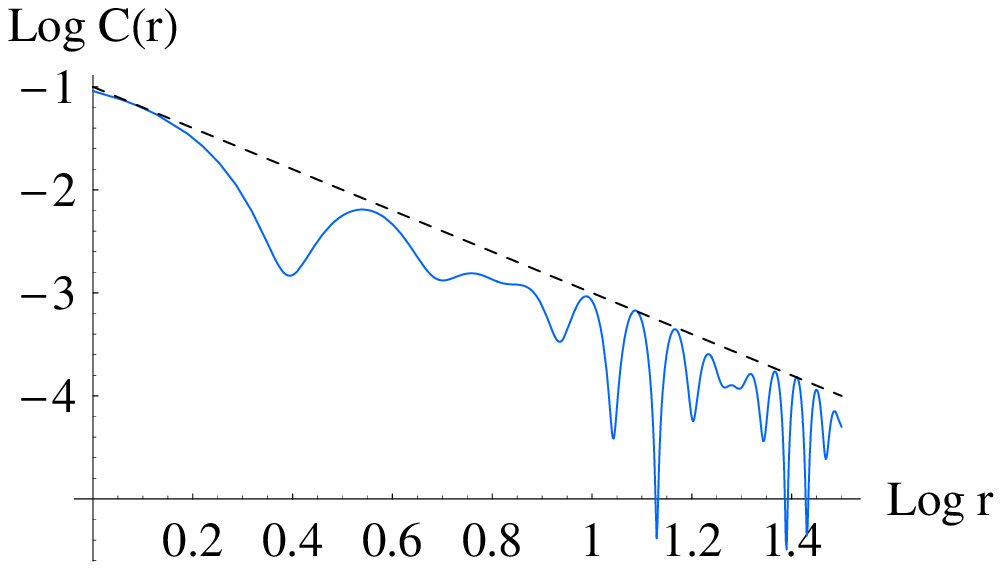}
\includegraphics[width=70mm,keepaspectratio,clip]{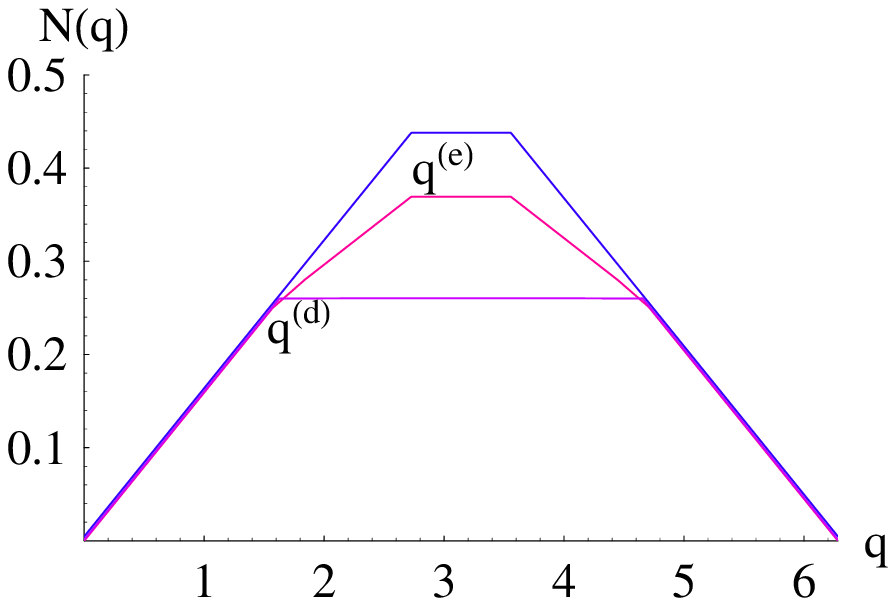}
\caption{(Color online) (a) Equal time density correlations $C(r)$ vs distance $r$ at $u=1$ and $t_x=0.6$; the dashed line represents the slope $1/r^2$. (b) $N(q)-d_{PS}$ at $t_x=.3$, $u=0$, and three different fillings: from top to bottom $n=.5$ (blue), $n=1$ (pink), $n=1.7$ (violet). }  \label{fig2}
\end{figure}

In order to clarify the meaning of the different modulations in the IPS phase, we also give in Fig. \ref{fig2} the charge-charge correlations $C(r) \doteq \langle ( n_i-n ) ( n_{i+r}-n) \rangle$ at half filling (upper part), and their Fourier transform $N(q)\doteq \sum_r e^{i q r} C(r)$ (lower part). Since $C(r)$ can be evaluated exactly (see for instance \cite{SCHAD}), we can also provide an analytic expression for $N(q)$, which reads:
\be
N(q)=C(0)+(\delta(q)-1) d_{PS} - a^2 \gamma_{k_F^{(e)}} (q)- (1-a^2) \gamma_{k_F^{(d)}} (q)\quad . \label{nq}
\ee
Here $\delta (q)$ is the Dirac delta, and $d_{PS}=(1-n+n_d') (n-1+n_e')$ is the constant responsible for the divergent contribution in $q=0$ characteristic of phase separation: at given ($u$, $t_x$) it becomes zero for $n<n_l\doteq 1-n_e'$, or $n>n_h\doteq 1+n_d'$. Moreover 
\be
\gamma_k(q)=-{1\over {4 \pi^2}} {\cal R} [ L_2 (e^{i (q+2 k)}) + L_2 (e^{i (q-2 k)}) - 2 L_2 (e^{i q}) ] \quad, \label{gamma}
\ee 
with $k=k_F^{(e)},k_F^{(d)}$, and $L_2 (z)$ denoting the dilogarithm of $z$ ({\cal R}[] being the real part of []), describes the $q$-dependence of density correlations. Equations (\ref{nq})--(\ref{gamma}) enlighten how in the IPS phase two wave-vectors do characterize $N(q)$, namely $q^{(e)}= 2 k_F^{(e)}$ and $q^{(d)}=2 k_F^{(d)}$, as shown in the lower part of Fig. \ref{fig2}.  Whereas in the CPS phase, since  $k_F^{(d)}=0$ and $\gamma_0 (q)=0$, only one feature characterizes $N(q)$, namely the wave-vector of the single fermion problem corresponding to a filling $n=1-n_e'$.
It can be recognized that the above type of ground-state phase diagram 
could explain different features displayed by the non-integrable 1D Hirsch model 
discussed in \cite{AAA}. At variance with the present model, in that case the model still had arbitrary $t_x$ ($t_x=2x -1$) but also $x'=0$, implying the appearance of a further term in the Hamiltonian. 
The numerical solution did show incommensurate charge-charge correlations 
(and superconductivity) at appropriate values of $u$ and $t_x$, always missing  
particle-hole invariance. Also, by suitably varying  
$u$ and $t_x$ at half-filling, the system could first pass from an insulating phase, 
characterized by flat density correlations, to a conducting phase in which 
that correlations acquire a modulation; then to the ICSS phase in which further 
modulations occur.  We suggest that the two latter phases could be identified with 
our  CPS and IPS phases respectively. As an example, in fig. \ref{fig2} we report 
$C(r)$ at $u$ and $t_x$ values identical with those used for the Hirsch model in fig. 3 of \cite{AAA}. The resemblance of the two figures is convincing.

In this paper, we have shown how the appearance of incommensurate characteristic wave-vectors  in strongly interacting fermionic  systems could be related to the occurrence of phase separation, in case the system is not symmetric under particle-hole transform. Recently the occurrence of incommensurate modulations in charge correlations was observed also in a Hubbard-like model including next-nearest neighbors hopping\cite{JNB}. Since that model is particle-hole asymmetric, we suggest that IPS could occur in that case. \\The mechanism could also be a natural candidate for explaining PS in unbalanced mixtures of cold fermionic atoms\cite{PAR,PIGI}. In this case the PI phase should be interpreted as the (unpolarized) superfluid phase, and the CPS phase should be identified with coexistence of the superfluid with the (polarized) Fermi gas.

The author acknowledges useful discussions with A. Anfossi, C. Degli Esposti Boschi, A. Andreev, B. Spivak; and the kind hospitality of the Department of Physics of the University of Washington for four months, during which this work was completed.


\begin{thebibliography}{10}
\bibitem{PAR} M.M. Parish et al., Nature Phys. 3, 124 (2007). 
\bibitem{PIGI} S. Pilati, S. Giorgini, Phys. Rev. Lett. {\bf 100} 030401 (2008)
\bibitem{XIal} G. Xiaolong et al., Phys. Rev. Lett. {\bf 98}, 030404 (2007) 
\bibitem{TRANQ} J.M. Tranquada et al., Nature {\bf 375}, 361 (1995)
\bibitem{EKL} V.J. Emery, S.A. Kivelson, and H. Q. Lin, Phys. Rev. Lett. {\bf 64}, 475 (1990) 
\bibitem{PLR} W.O. Putikka, M.U. Luchini, and T.M. Rice, Phys. Rev. Lett. 68, 538  (1992)
\bibitem{HEMA} C. S. Hellberg, and E. Manousakis, Phys. Rev. Lett. {\bf 78}, 4609 (1997)
\bibitem{CBS} M. Calandra, F. Becca, and S. Sorella, Phys. Rev. Lett. {\bf 81}, 5185 (1998) 
\bibitem{cam}
J. T. Gammel and D. K. Campbell, Phys. Rev. B \textbf{60}, 71 (1988);
Y. Z. Zhang, \emph{ibid} \textbf{92}, 246404 (2004);
\bibitem{SA} M.A. Simon, and A.A. Aligia, Phys. Rev. B 53, 15327 (1996)
\bibitem{corrhop} F.H. Essler, V. Korepin, and K. Schoutens, Phys. Rev.
Lett. {\bf 68}, 2960 (1992); Phys. Rev. Lett. {\bf 70}, 73 (1993); R.Z. Bariev et al., J Phys A {\bf 26}, 1249 (1993); A. Montorsi, D.K. Campbell, Phys. Rev. B {\bf 53}, 5153 (1998) 
\bibitem{hirsch}
J. E. Hirsch, Physica C \textbf{158} 326 (1989); J. E. Hirsch and
F. Marsiglio, Phys. Rev. B \textbf{39}, 11515 (1989).
\bibitem{JAKA} G.I. Japaridze, and A.P. Kampf, Phys. Rev. B 59, 12822 - 12829 (1999)
\bibitem{NAKA}  M. Nakamura, K. Itoh, N. Muramoto, J. Phys. Soc. Jpn. 70, 3606 (2001); see also M. Nakamura, T. Okano and K. Itoh, Phys. Rev. B 72, 115121
(2005)   
\bibitem{ADMO}A. Anfossi, C. Degli Esposti Boschi, A. Montorsi, and F. Ortolani,
Phys. Rev. B \textbf{73}, 085113 (2006).
\bibitem{AAA} A. Aligia {\it et al.}, Phys. Rev. Lett. \textbf{99}, 206401 (2007) 

\bibitem{AAS}L. Arrachea and A. A. Aligia, Phys. Rev. Lett. \textbf{73}, 2240 (1994);
J. de Boer, V. E. Korepin, and A. Schadschneider, \emph{ibid} \textbf{74}, 789 (1995).
\bibitem{PEMI} K. Penc, and F. Mila, Phys. Rev B {\bf 49}, 9670 (1994).
\bibitem{SCHAD} A. Schadschneider, Phys. Rev. B {\bf 51}, 10386 (1995); A. Anfossi, P. Giorda, and A. Montorsi, Phys. Rev. B \textbf{75}, 1 (2007).
\bibitem{JNB} G. I. Japaridze, R. M. Noack, and D. Baeriswyl, Phys. Rev. B \textbf{76}, 115118 (2007). 

\end{thebibliography}
\end{document}